\documentclass[a4]{IEEEtran}
\usepackage{cite}
\usepackage{amsmath,amssymb,amsfonts}
\usepackage{algorithmic}
\usepackage{graphicx}
\usepackage{textcomp}
\usepackage{xcolor}
\usepackage{multirow, booktabs}
\usepackage{listings,multicol}
\usepackage{xfrac}
\usepackage[utf8]{inputenc}
\usepackage[T1]{fontenc}
\usepackage{fixltx2e}
\usepackage{csquotes}
\usepackage{amsmath,diagbox,tabu}
\usepackage{hyperref}
\usepackage{comment}

\usepackage{siunitx}
\usepackage{pgfplots}

\usepackage{todonotes}

\usepackage{url}

\begin{document}

\title{TRUSTD: Combat Fake Content using Blockchain and Collective Signature Technologies}

\author{
    \IEEEauthorblockN{\IEEEauthorrefmark{1}, Mohamad Alissa\IEEEauthorrefmark{1}, William J Buchanan\IEEEauthorrefmark{1}}
    \IEEEauthorblockA
    {\IEEEauthorrefmark{1}
    \\z.Jaroucheh@napier.ac.uk, w.buchanan@napier.ac.uk}

}

\author{\IEEEauthorblockN{1\textsuperscript{st} Zakwan Jaroucheh}
\IEEEauthorblockA{\textit{Blockpass ID Lab} \\
\textit{Edinburgh Napier University}\\
Edinburgh\\
z.jaroucheh@napier.ac.uk\\}
\and
\IEEEauthorblockN{2\textsuperscript{nd} Mohamad Alissa}
\IEEEauthorblockA{\textit{Blockpass ID Lab} \\
\textit{Edinburgh Napier University}\\
Edinburgh\\
m.Alissa@napier.ac.uk\\}
\and
\IEEEauthorblockN{3\textsuperscript{rd} William J Buchanan}
\IEEEauthorblockA{\textit{Blockpass ID Lab} \\
\textit{Edinburgh Napier University}\\
Edinburgh\\
w.buchanan@napier.ac.uk}

}

\maketitle
\begin{abstract}
The growing trend of sharing news/contents, through social media platforms and the World Wide Web has been seen to impact our perception of the truth, altering our views about politics, economics, relationships, needs and wants. This is because of the growing spread of misinformation and disinformation intentionally or unintentionally by individuals and organizations. This trend has grave political, social, ethical, and privacy implications for society due to 1) the rapid developments in the field of Machine Learning (ML) and Deep Learning (DL) algorithms in creating realistic-looking yet fake digital content (such as text, images, and videos), 2) the ability to customize the content feeds and to create a polarized so-called “filter-bubbles” leveraging the availability of the big-data. Therefore, there is an ethical need to combat the flow of fake content. This paper attempts to resolves some of the aspects of this combat by presenting a high-level overview of TRUSTD, a blockchain and collective signature based ecosystem to help content creators in getting their content backed by the community, and to help users judge on the credibility and correctness of these contents.

\end{abstract}

\IEEEpeerreviewmaketitle
\section{Introduction}
Fake news has become increasingly one of the main threats to democracy, journalism, and freedom of expression. It has weakened public trust in governments and its potential impact on the contentious \emph{Brexit} referendum and the equally divisive 2016 U.S. presidential election is yet to be realized \cite{survey2}. Our economies are also affected by the spread of fake news, with fake news being connected to stock market fluctuations and massive trades. For example, fake news claiming that Barack Obama was injured in an explosion wiped out \$130 billion in stock value \footnote{https://www.forbes.com/sites/kenrapoza/2017/02/26/can-fake-news-impact-the-stock-market}. 

The news is a newly received or noteworthy information, especially about recent not-necessarily political events. Deception can be described as an act of "intentionally causing another person to have or continue to have a false belief that is truly believed to be false by the person intentionally causing the false belief by bringing about evidence on the basis of which the other person has or continues to have that false belief" \cite{james}. With the ease of sharing information on social media platforms, the rapid accessibility of uploaded content on the World Wide Web, and the rapidly progressing fields of Artificial Intelligence (AI) and Machine Learning (ML), deception has far more potential in altering an individual’s perception and influencing their decisions and choices than it has ever had. In this context, various issues can be identified:

\textbf{1 - Fake content can have a strong impact on listeners and readers:} Experts distinguish between misinformation and disinformation \cite{oncombating}. Misinformation is a false or misleading piece of information —e.g., sharing a fraudulent content without verifying the authenticity of its source is misinformation. Disinformation is deliberately falsified information to obscure the truth —e.g., spreading false content with the intention to harm an individual’s reputation. In this paper we refer to the misinformation, disinformation and artificially generated content as deceptive artifacts or fake content. Fake content is a general term which would include the term fake news. Throughout the paper, we will use the terms fake news and fake content interchangeably. Repeated exposure to a piece of information makes it familiar, until it is eventually perceived as acceptable and valid. This phenomenon is known as the "illusory truth" effect \cite{james}.

\textbf{2- Individuals are victims of the "filter bubbles":} Social media websites, including Facebook, Twitter and LinkedIn are said to construct “filter bubbles” that allows users to only view content they agree with or that aligns with their pre-existing beliefs \cite{bubble}. A filter bubble is an individual’s personal and unique online space, the nature of which depends on their online identity (e.g. preferences, behaviour, believes, etc.). Individuals may not have full control on the construction of this identity. In addition, individuals have no say regarding what penetrates into the space and what gets filtered out. Therefore, individuals are now victims of the algorithms that these centralised platforms use to filter and sort the content (content feed). There is no guarantee that these algorithms are not designed to favor one content over another in order to maximise the revenue or achieve a specific objective (e.g. Cambridge Analytica scandal). 

\textbf{3- Lack of transparency and traceability:} Anyone has the freedom to publish any content on any social media platform without any scrutiny about the correctness or the credibility of the content. Individuals do not have any mechanism to trace back the originator of the content. In addition, if the content is published in a news agency Website, individuals do not have any clue who has approved that content and what is the approval process.

\textbf{4- Lack of empowering tools:} There is a debate whether the centralised social media companies should intervene in controlling the political adverts on their platforms or not. Some people argue in favour of banning these adverts, while others argue that these private companies should not make the decision on behalf of the user and these adverts should be shown to the user, and it is up to the user to make their mind. We have two issues with this argument: 1) since it is an advertisement platform, their algorithm could be designed to allow these adverts to reach to the users in a way that maximises the profit and maybe maximises the intended impact on the users, 2) there is no helpful tools for the users to assess the credibility and the correctness of the content. The users are making their own judgement on any content in an ad-hoc manner. Verifying the authenticity and credibility of every content is a daunting task for the users. That is why some users prefer to trust a few sources of news/information.

In order to remedy to the above issues we propose TRUSTD, an ecosystem that allows content readers (users) to assess the correctness and credibility of the content they receive. We believe that there is a need to empower the user with a tool that allows them to delegate the task of verifying the content to a set of trusted parties of their own choice. That way, users will be able to assess the credibility of the content themselves based on their level of trustiness of each of their trusted parties. In addition, since this is a global issue, TRUSTD is a decentralised and open platform where everyone is able to join the system. 

Today, Distributed Ledger Technologies (DLTs) and specifically blockchain, the Decentralised Identifiers (DIDs) \cite{did}, and collective signature \cite{syta2016keeping} present opportunities for stakeholders and policymakers as potential technologies that can help to combat digital deception. These technologies enable security and trust in a decentralised Peer-to-Peer (P2P) network without any central managing authority. There are only a few articles of the literature that use blockchain to combat digital deception and they are mostly focused on tracing the source of the information. To the knowledge of the authors, this is the first article that proposes a user-centred approach to empower the user with the a tool to identify digital deception using the above technologies. In this paper, we are focusing on the human element when addressing the fake content issue. In other words, it is true that the machines are able to detect the fake content to some extent but there is no replacement to the human intervention. In addition, TRUSTD does not inhibit the current existing automated tools to detect the fake content, these deep tools can be included as part of the user policy rather than rely only on the quality of these sophisticated tools.

The rest of this paper is organized as follows. Section II reviews the state-of-the-art of current digital deception and the involved technologies and approaches used to combat it. Section III provides a background on the Schnorr signing and how it is used in the collective signing algorithm. In Section IV, we list the main driving requirements that drive the design of TRUSTD. Section V describes the details of the proposed architecture and its implementation. Finally, Section VI is devoted to conclusions and future work.

\section{Related Work}
Recently, Wikipedia co-founder Jimmy Wales has launched "WikiTribune"\footnote{https://wt.social/}, a platform to combat the low-quality content. This platform is designed for small, niche communities that can sustain themselves where almost everything on the platform is editable (similar to Wikipedia) and the users are responsible to correct the contents even those written by other users. However, this might not be an effective approach to combat the fake content as this platform can be turned into a source of fake news. 

Governments worldwide are taking various measures to prevent the widespread of fake news, which may be driven by different motivations and which may undermine national security \cite{survey1}. These measures include the introduction of new legislation e.g. 1) new laws that would give governments more powers to hold technology companies (e.g., Facebook, Twitter and Google) and individuals accountable for the spread of fake news, and 2) new laws that would seek to counter the impact of automated social media accounts (bots). Although these laws are important and in the right direction, they hold the technology companies accountable of identifying the fake news on behalf of the content readers (users). In this respect, users have to trust that these companies implement adequate algorithms to remove the fake news as quickly as possible. However, this is not enough because the user has to trust the criteria followed by these algorithms to identify the fake news. The user should be empowered by tools to allow them to specify their trust policy in order to determine the credibility of the content. 

In response, technology companies are defending themselves and are introducing mechanisms to detect and remove fake news. For example, Facebook Journalism Project aims to collaborate with content organizations and journalism experts to improve the quality of information shared on the platform \cite{facebook1}. Facebook tries to provide an improved ranking system for posts shown in the News Feed and expediting the reporting of misleading content. Google has presented a white paper outlining measures to prevent the spread of disinformation through their products \cite{google}. The objective is to enhance its search result ranking system by ranking news on the basis of expertise, credibility and authority. Although these initiatives are necessary, they still do not position the user in the center. The user should be able to make their own judgement on the content based on their own criteria. We believe the fake news problem cannot be solved by automated tools only; and these tools should be complemented by manual verification process conducted by humans. 

The IEEE Global Initiative has published a report titled “Ethically Aligned Design” \cite{ieee-initative} on guiding and encouraging the development of autonomous systems that are primarily focused on human well-being and protecting human rights through preventing misuse of AI, ensuring system transparency and developing a framework for developer accountability. The goal of The IEEE Global Initiative is that Ethically Aligned Design will provide pragmatic and directional insights and recommendations, serving as a key reference for the work of technologists, educators and policymakers in the coming years. TRUSTD is inspired by the general principals of the above initiative such as human rights, well-being, data agency, transparency and accountability; and it uses these principals as a guideline to derive the driving requirements in the design of an open and decentralised ecosystem to combat the flow of fake content. 

The authors in \cite{survey2}, presented four perspectives to study fake content: knowledge-based, style-based, propagation-based and credibility based. In knowledge-based approach, one aims to analyze and/or detect fake news, using a process known as fact-checking. Manual fact-checking can be divided into (I) expert-based and (II) crowd-sourced fact-checking. Expert-based fact-checking relies on domain-experts to verify the given news contents, therefore, it leads to highly accurate results. Recently, many websites have emerged to provide expert-based fact-checking services. For example, PolitiFact provides “the PolitiFact scorecard”, which presents statistics on the authenticity distribution of all the statements related to a specific topic. Another example is HoaxSlayer, which classifies the articles and messages into e.g., hoaxes, spams and fake news. A comprehensive list of fact-checking websites is provided by Reporters Lab at Duke University\footnote{https://reporterslab.org/fact-checking/}, where over two hundred fact-checking websites across countries and languages are listed. Although these websites can provide ground-truth for the detection of fake content, these websites are still silo-ed and centralised and they have their own "experts" and their own methodology of identifying the fake news. 

Crowd-sourced fact-checking relies on a large number of fact-checking individuals and therefore, it is less credible and accurate due to the political bias of these individuals and their conflicting annotations. Hence, one often needs to (i) filter non-credible individuals and (ii) resolve conflicting fact-checking results. An example is Fiskkit\footnote{https://fiskkit.com/}, where users can upload articles, and provide ratings and tags for sentences within articles. TRUSTD can be seen as a crowd-sourced fact-checking approach but it is the content creator who chooses the fact-checkers and it is the user who determines the content's credibility based on their trust level in these fact-checkers.  

Automatic fact-checking techniques have been developed, heavily relying on Information Retrieval (IR) and Natural Language Processing (NLP) techniques. The overall automatic fact-checking process can be divided into two stages: (I) fact extraction (also known as knowledge-base construction) and (II) fact-checking (also known as knowledge comparison) \cite{survey2}. Knowledge-based approaches aim to evaluate the authenticity of the given content, while style-based approaches aim to assess the intention behind publishing the content. Credibility-based approaches evaluate the content based on content-related and social-related information. The AI Foundation has developed an intelligent software called “Reality Defender”, to detect potentially fake media in the digital world\footnote{https://aifoundation.com/responsibility/}. This software runs AI-driven analysis techniques to detect alterations in the scanned images, videos and other media, and allows for reporting suspected fakes. In TRUSTD, these deep tools can be included as part of the user's trust policy, as will be seen later, rather than rely solely on the quality of these tools.

Blockchain technology has been already leveraged to contribute solutions to address the fake news problem. For example, Shang et al. \cite{Shang} trace the source of news by keeping a ledger of timestamps and the connections between the different blocks. Huckle et al. \cite{bc1} introduced a blockchain-based application that is capable of indicating the authenticity of digital media. Using the trust mechanisms of blockchain technology, the tool can show the provenance of any source of digital media, including images used out of context in attempts to mislead. However, the authors mentioned that although their application has the potential to be able to verify the originality of media resources, that technology is only capable of providing a partial solution to fake content. We agree with the authors that this is because the blockchain technology is incapable of proving the authenticity of a content story and that requires human skills. 

\section{Background}
\subsection{Schnorr signing}
In Schnorr signing, we can aggregate public keys of $P$ participants into a single signing key \cite{ohta1999multi}, and uses the non-interactive version of the Fiat-Shamir heuristic \footnote{https://en.wikipedia.org/wiki/Fiat-Shamir\_heuristic}.  Using Elliptic Curve methods, to sign a message we take a random value ($k$) and a private key value ($d$) and a generator point $G$ ($G$ is a base point on an elliptic curve) and compute:

\begin{equation}
Q=dG
\end{equation}

and:

\begin{equation}
R=kG
\end{equation}

For it to be non-interactive we then calculate:
\begin{equation}
e=H(R\parallel M)
\end{equation}

and then:
\begin{equation}
s=k-ed
\end{equation}

The signature is then $(s,e)$. To verify we compute:
\begin{equation}
r_{v}=sG+eQ    
\end{equation}

and then:

\begin{equation}
e_{v}=H(r_{v}\parallel M)
\end{equation}

We then check that $r_{v}=e_{v}$.


Each participant has a private key ($a_i$) and a public key $A_i=a_i G$. We can then determine the aggregate public key with: 

\begin{equation}
A = \sum_{i \in P} A_i
\end{equation}

\subsection{Collective Signing}
Within legal infrastructures, we might have several witnesses $W$, and we ask a number of the witnesses $W’$ to verify that something is correct. If one of the witnesses cannot verity the information, we would highlight a problem. Let us say we have a controller on a network, and a number of selected trusted witnesses. Each of the witnesses can then check all of the messages sent by the nodes on the network, and if one of them determines a problem, they can tell the rest of the network. In this respect, every message ($M$) is collectively signed by $W$ witnesses. 

The collective signing (CoSi) algorithm is defined by Syta et al \cite{syta2016keeping}. With CoSi (collective signing), there are four phases involving $P$ participants and where the leader has an index value of zero. Each participant has a private key ($a_i$) and a public key ($A_i=a_iG$, and where $G$ is a base point on an elliptic curve). We then determine the aggregated public key with \cite{syta2016keeping}:

\begin{equation}
A = \sum_{i \in P} A_i
\end{equation}

\textbf{Announcement:} Initially the leader broadcasts a message ($M$) that it wants the participants to sign.

\textbf{Commitment:} Each node $i$ will pick a random scalar ($v_i$) and determines their commitment ($V_i=[v_i]G$). Each commitment is then sent to the leader, who will wait for a specific amount of commitments ($P’$) to be received. The leader then creates a participant bitmask and aggregates all the received commitments:

\begin{equation}
V = \sum_{j \in P'} V_j
\end{equation}

and creates a participation bitmask $Z$. The leader then broadcasts $V$ and $Z$ to the other participants.

\textbf{Challenge:} Each of the participants computes the collective challenge (using the hash function $H$):

\begin{equation}
c = H(V || A || M)
\end{equation}

and send the following back to the leader:

\begin{equation}
r_i = v_i + c \times a_i
\end{equation}

\textbf{Response:} The leader will wait until the participants in $P’$ have sent their responses. Once received, the leader computes the aggregated response:

\begin{equation}
r = \sum_{j \in P'} r_j
\end{equation}

and publishes the signature of the message (M) as:
\begin{equation}
(V,r,Z)
\end{equation}

Each node can then check their own signature value and agree with the leader. 

In order to handle large numbers of participants during signature generation efficiently, CoSi protocol uses a tree-shaped network communication overlay \cite{cosi-ietf}. Any tree used by CoSi should be a complete tree for performance reasons. The leader is the root node of the tree and is responsible for creating the tree. An intermediate node is a node who has one parent node and at least one child node. A leaf node is a node who has only one parent and no child nodes. The leader multicasts a message to his direct child nodes. Upon reception of a message, each node stores the message and multicasts it further down to its children node, except if the node is a leaf. Each node generates its response. Each leaf node sends its response to their parent and is allowed to leave the protocol. Each other node starts waiting for the responses of its children. When the root node receives all the responses from its children, it can generate the signature.

\subsection{Decentralised Identifier (DID)}
The self-sovereign identity (SSI) refers to an identity management system which allows individuals to fully own and manage their digital identity \cite{ssi}. The World Wide Web Consortium (W3C) working group on verifiable claims states that in a SSI system users exist independently from services \cite{vc}. This highlights the contrast to current identity management which either relies on a number of large identity providers such as Facebook and Google or the user has to create new digital identities at each individual service provider. SSI is enabled by the new development of blockchain technology. Through the trustless, decentralised database that is provided by a blockchain, classic Identity Management registration processes can be replaced \cite{ssi}.

A Decentralised Identifier (DID) is comprised of a scheme as well as a method and method specific identifier. The method closely resembles the namespace component of an Uniform Resource Name (URN) \footnote{\url{https://en.wikipedia.org/wiki/Uniform_Resource_Name}}. Each distinct blockchain or rather each identity registry constitutes its own namespace while the blockchain specific identifiers specify
the actual identity addressed by the DID. An example for such a DID path would be: did:examplechain:123456789. The DID document is the key to the decentralised identity.

A DID is a unique identifier that can be resolved to a DID document \cite{did_spec}. This document contains cryptographic material and authentication suites enabling identification of the DID subject. They also contain service endpoints allowing secure encrypted communication to the DID subject. A public DID is a DID that is registered on a public distributed ledger meaning DID is resolvable to a document and hence verifiable by anyone. There should be some restrictions or process by which DIDs can be registered to a public ledger. Once registered it is possible to use that DID as a root of trust.

Establishing pairwise private DID connections can be done by first contacting the endpoint (i.e. agent) resolved from the public DID document. Sending a new unique DID and DID document along with proof you are the DID subject. In response you should receive a new unique DID signed by the public key of the public DID. Together these two DIDs form a private unlinkable connection between two parties complete with private endpoints and the cryptographic material required to verify the origin and integrity of the communication.

\subsection{DLTs and blockchain capabilities}
DLTs like Tangle or blockchain are able to provide seamless authentication, efficient and secure data storage, robustness against attacks, scalability, transparency and accountability. Such features can play an effective role in combating fake content, considering that transactions cannot be tampered once they have been distributed, accepted and validated by a network consensus and stored in blocks \cite{bc2}. Moreover, transactions are easily auditable by all the involved stakeholders. More detailed information on how to design a blockchain according to the business needs and deployment environment can be found, for example, in \cite{bc3}\cite{bc4}.

\section{Driving Requirements}
We believe that combating the fake content cannot be a pure technical solution and that a human element should always exist. In the proposed ecosystem, we can identify two types of users: 

- Individuals and organizations who produce the content. We call them content creators (CCs). These creators may have different agendas and they could be honest or dishonest in reporting any news or spreading any information.  

- Content readers: We call them users. These users are the consumers of the content. 

The following are the set of requirements that we believe TRUSTD needs to meet: 

\subsection{Traceability of origin}
The ecosystem should allow the originator of the content to be identified at any point of time. That could be achieved by storing the content along with the originator ID in an immutable ledger that cannot be maliciously altered (i.e. blockchain). 

\subsection{Traceability of approval}
In TRUSTD, every CC can request other actors to "sign" their content. This "signature" means that the signing actor agrees on the correctness of the content. This is necessary in order for the user to be able to assess the credibility of the content, as will be seen in the next sections. Once any actor provides their signature (witnessing the correctness of the content), this signature should be stored in a way that this actor cannot deny their action at any time. However, the ecosystem should allow any actor to change their mind and revoke their signature after, for example, they discover that the content was incorrect. For transparency purposes, the list of actions of any actor should be captured in the immutable ledger.

\subsection{Accountability}
There is a need to ensure that the signing actors cannot deny their involvement in signing off the content when the content is discovered to be false.  

\subsection{System openness}
The ecosystem should allow any individual or organization to join and be part of the approval actors. For example, these individuals can be journalists, editors, activists, etc. who have different levels of expertise; and the organizations can be any news agency. However, since the fake content concerns all people and it affects everyone's life, the system should be open to allow anyone to be part of the signing actors - we call them appraiser actors (AAs). This is especially important for the content that is originated from people (non-journalists) such as abnormal accidents or natural disasters, and that the only witnesses are the people seen at the accident location.

\subsection{The content should be evaluated by the user}
It is true that the ecosystem allows the invited AAs to sign off the contents they receive from CCs. However, the system should keep the user in the center and it should empower them with a mechanism to judge on the credibility of the content. For that objective, the user should be able to select the list of AAs and assign them a trust level value which is in the range $[0, T]$. $T$ is the maximum trustiness value. For example, a trust level of $T$ means the user completely trusts that actor. A trust level of 0 means the user does not trust the actor. 

The user can determines the credibility of the content by calculating the following formula:

\begin{equation}
C = \sum_{i \in N}T_i/(T*N) 
\end{equation}

$Ti$: The trustiness value of the actor $i$.
$N$: The length of the actor list added by the user.
The result is a number in the range $[0, 1]$. This number indicates the level of credibility of the content. 

\subsection{The content should be backed by actors}
Before publishing their content to the social media and news websites, the CCs can get an "approval" from a set of AAs of their choice. These AAs can be selected by the CCs themselves or by the system in automated way based on a specific criteria. For example, the system can extract the main keywords of the content and send it to all AAs that are tagged with these keywords. In addition, these AAs can specify a policy (using the DID system) to accept signature requests only from CCs who can prove that they have specific attributes, for example, holding a specific degree. These attributes should be signed by an entity trusted by the AA. 

The approval comes in a form of a collection of signatures from these AAs indicating that they agree on the correctness of the content. We can imagine here different signature types. For example, the CC may obtain a collection of signatures from the AAs who share the CC's opinion, or who support the CC's call. In this paper, we focus on the signature that means that the AA agree on the correctness of the content. 

\subsection{The ecosystem should be decentralised and scalable}
Since the fake news/content is a global issue, the system should not be centralised. Instead, the system should provide a protocol between the different users and actors which are registered in different subsystems owned by different companies or organisations. However, these subsystems should have one common denominator: the identifier of the users and actors. For that aim, the decentralised identity (DID) \cite{did} would be a perfect fit in this context. In this respect, the actors can create their own DIDs and publish them on their websites or their social media profiles. That DID will be used for signature verification as will be seen later. 

\begin{figure*}
  \includegraphics[width=\textwidth]{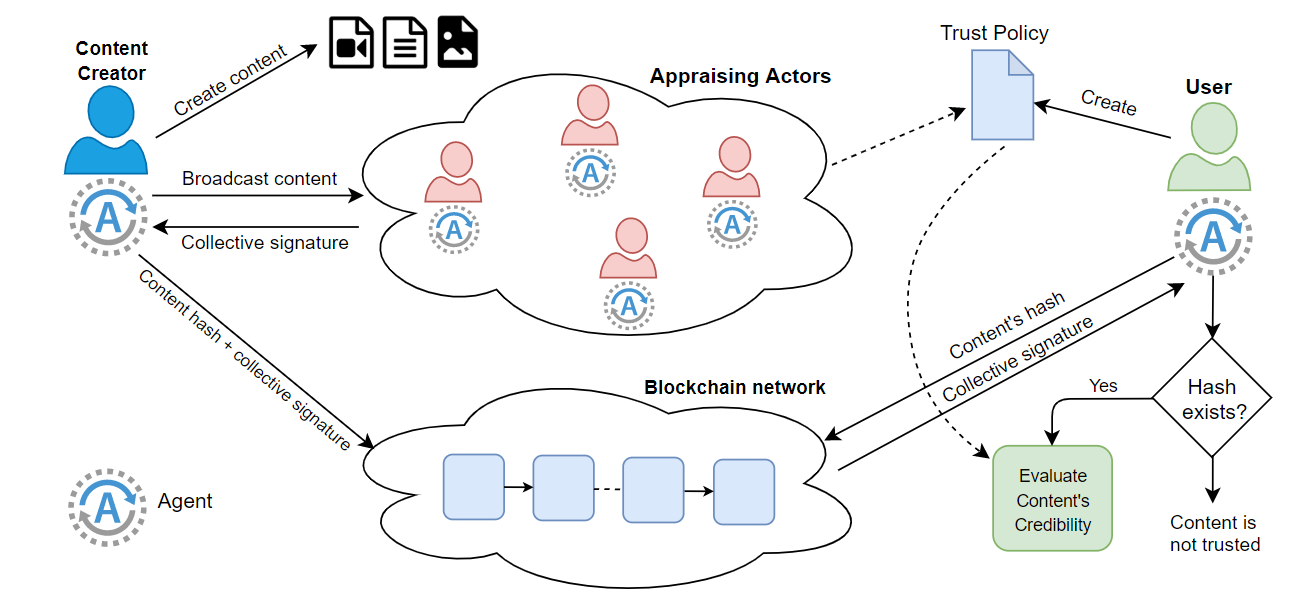}
  \caption{The TRUSTD approach}
  \label{fig:fig1}
\end{figure*}

\section{System Design}
Three main entities can be identified: 1) The content creator (CC), 2) the appraisal actor (AA) who signs off any content, and 3) the user who needs to assess the credibility of any content. Each entity should be represented by a DID agent. These agents should follow a specific protocol in order to get any content signed off by the relevant AAs. The users' agents should be able to verify any content based on the trust policy specified by the user. Here is the proposed protocol as illustrated in Fig 1:

1- An AA creates its own DID and publishes it somewhere so that the CCs, users and other AAs know about it.

2- A CC creates the content in a form of a text, photo, video, sound, or any other digital format.

3- Depending on the content scope, the CC can choose a list of AAs that they are candidate to approve and sign off the content. Alternatively the system recommends specific AAs. The result is a set of DIDs which can be used to retrieve their corresponding public keys. 

4- A collective signature is generated in five steps over two round trips between the agent of the CC and the agents of the selected AAs, as follows:

AAs need not coordinate for the creation of their key-pairs beyond selecting a common elliptic curve, and verifiers can apply flexible acceptance policies beyond simple t-of-n thresholds. 

We use the following notation: 

$B$: Generator of the group of AAs.

$L$: Order of the group generated by B.

($a_i$, $A_i$): Each AA$_i$ generates their long term private-public key pair ($a_i$, $A_i$) as in EdDSA.

$A$: collective public key $A$ generated from the public keys of AAs.

$N$: denotes the list of AAs, the size of $N$ is denoted by $n$.

- \textbf{Announcement Step:} The CC broadcasts an announcement message to the AAs indicating the start of a signing process. This message contains the content itself.

- \textbf{Commitment Step:} Upon the receipt of the announcement message, each AA$_i$ generates a random secret $r_i$ by hashing 32 bytes of cryptographically secure random data. Each $r_i$ must be re-generated until it is different from ($0\ mod\ L$) or ($1\ mod\ L$). Each AA then constructs the commitment $R_i$ as the encoding of $[r_i]B$, sends $R_i$ to the CC and stores the generated $r_i$ for usage in the response phase. 

- \textbf{Challenge Step:} The CC waits to receive the commitments $R_i$ from the other AAs for a certain configurable time frame. After the timeout, the CC constructs the subset $M$ of AAs from whom he has received a commitment $Ri$ and computes the sum: 

\begin{equation}
R = \sum_{i \in M}R_i
\end{equation}

The CC then computes 
 
\begin{equation}
c = SHA512(R || A || M) \ mod\ L
\end{equation}
The CC broadcasts $c$ to all AAs.

- \textbf{Response Step:} Upon reception of $c$, each AA generates their response:

\begin{equation}
s_i = (r_i + c * a_i)\ mod\ L
\end{equation}

 and send it to the CC. 

- \textbf{Signature Generation Step:} The CC waits to receive the responses $s_i$ from the AAs for a certain configurable time frame. After the timeout, the CC checks if he received responses from all AAs in $M$ and if not he must abort the protocol. The CC then computes the aggregate response 

\begin{equation}
s = \sum_{i \in M}s_i\ mod\ L
\end{equation}

and initializes a bitmask $Z$ of size $n$ to all zero. For each AA$_i$ who is present in $N$ but not in $M$ the CC sets the $i-th$ bit of $Z$ to $1$, i.e., $Z[i] = 1$. The CC then forms the signature $sig$ as the concatenation of the byte-encoded point $R$, the byte-encoded scalar $s$, and the bitmask $Z$. The resulting signature is of the form: 
 
\begin{equation}
sig = R || s || Z
\end{equation}
 
5- The CC sends a request to the underlying blockchain to store the hash of the content along with the collective signature in the blockchain. 

6- In order to verify the credibility of a specific content, the user agent calculates the hash of the content and uses it as an input parameter to query the blockchain and to retrieve the corresponding collective signature (if any). 

7- As aforementioned, each user should be able to choose a set of their trusted AAs. Each user agent maintains a list of the public keys of those actors along with their trust level values. Based on the retrieved collective signature, the user agent obtains the public keys of the AAs who signed off the content and then calculates the credibility value of the content as described in section IV.E.  

\section{Implementation}
\begin{figure}[p]
\centering
\includegraphics[keepaspectratio,height=3in, width=3in]{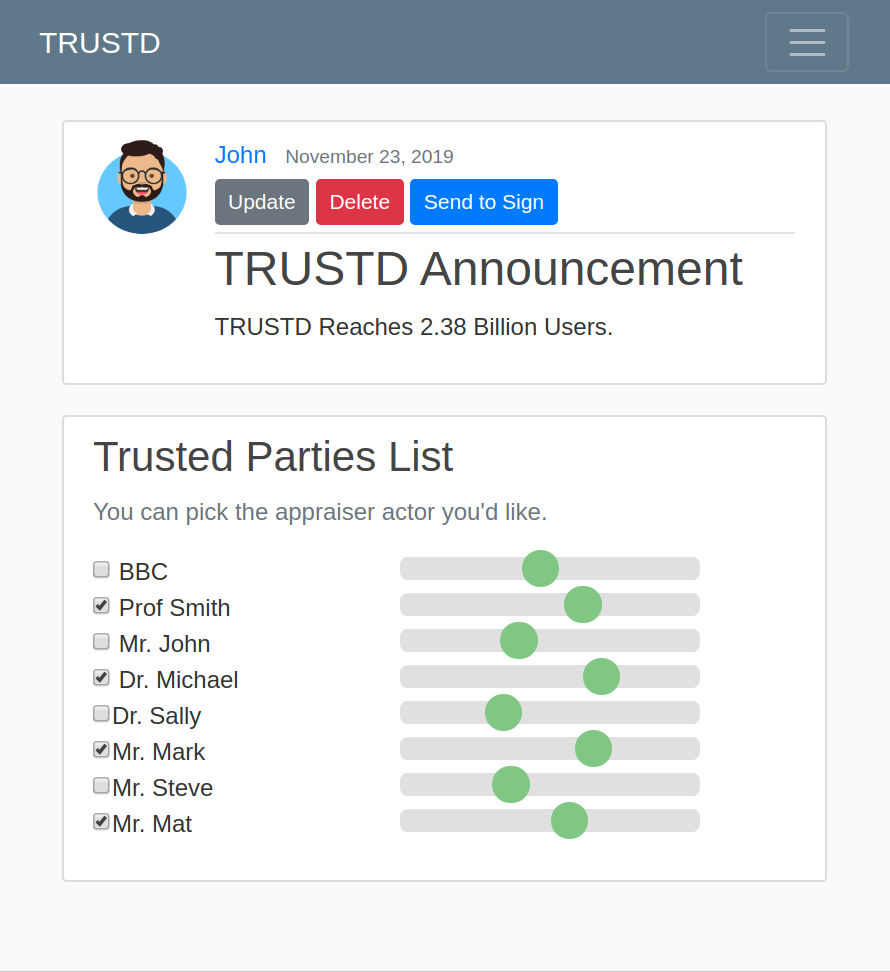}
\caption{The user assigns a trust level to each trusted party, and sends the content to a subset of them}
\label{fig:User_View_Send}
\end{figure}

\begin{figure}[p]
\centering
\includegraphics[keepaspectratio,height=3in, width=3in]{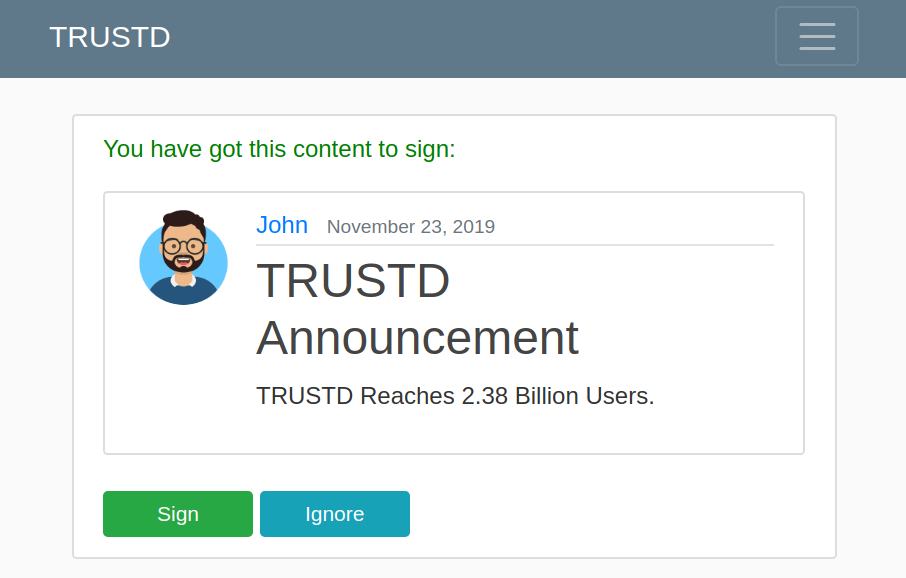}
\caption{The user receives requests from other users to sign contents}
\label{fig:User_View_Sign}
\end{figure}

\begin{figure}[p]
\centering
\includegraphics[keepaspectratio,height=3in, width=3in]{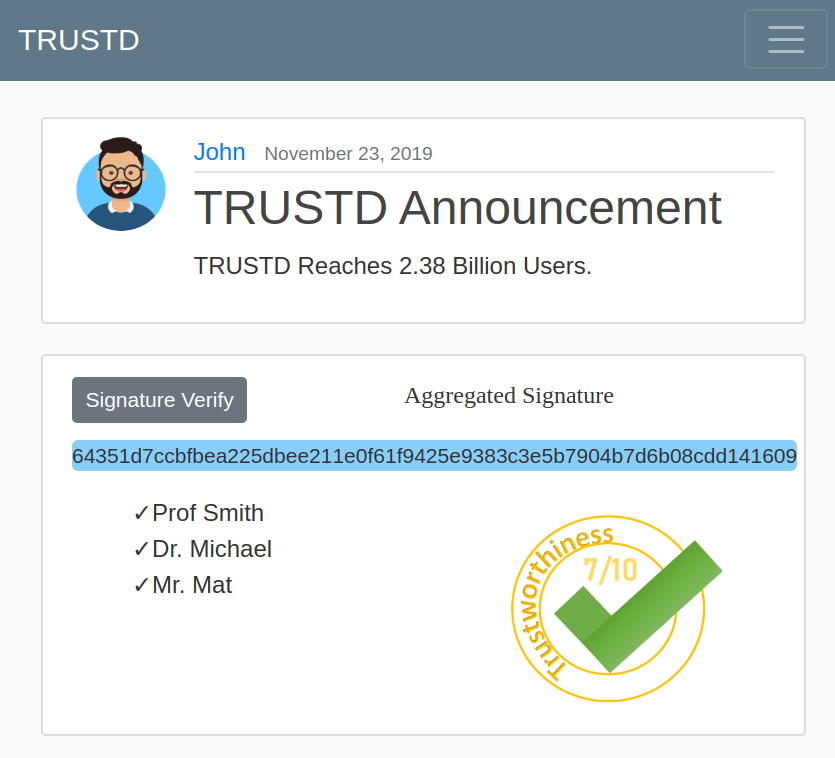}
\caption{The user can see the names of parties who signed the content, along with the trustworthiness level of the content}
\label{fig:User_View_Recv}
\end{figure}

An initial prototype has been built (Fig. \ref{fig:User_View_Send} to \ref{fig:User_View_Recv}). We use Django version 2.2.7 to build a TRUSTD-based blog where the users can register to the system, create contents, specify the list of their trusted parties, assign a trust level to each actor in that trusted list, send the contents to a subset of their trusted parties list (Fig \ref{fig:User_View_Send}), view the names of the signed parties along with the trustworthiness level of the related content (Fig \ref{fig:User_View_Recv}), and receive requests from other users to sign contents (Fig \ref{fig:User_View_Sign}). We use GO programming language version 1.13.4 to implement CoSi\footnote{\url{https://asecuritysite.com/encryption/go_cosi}}. 

\section{Discussion}
\subsection{Limitation of the TRUSTD approach}
In its current implementation, TRUSTD has some limitations that can be solved in future work:

- The authors in \cite{OnTheSecurity} introduced mBCJ, a secure two-round multi-signature scheme. Their results show that mBCJ is only marginally less efficient than CoSi, so that any protocol based on the CoSi scheme should instead be built on the provably more secure mBCJ scheme. The TRUSTD approach is agnostic to the used collective signature scheme. We plan to adopt the mBCJ scheme as a mechanism to collectively sign any digital content.

- In some extreme cases, the content could be signed by one AA who is not trusted by the user. For example, if the content is a video of a publicly known person and this person signs off this video as correct, but this person is not trusted by the user, in our current implementation, the system recommends to the user that the video is not credible even though that AA is the only AA who can sign off this video. Therefore, we need to extend the way we calculate the credibility of any content to consider different dimensions, each of which takes two extremes, for example: [liar, truthful], [enemy, friend], [dishonest, honest], etc.

- In order for the CC to get their content signed off by the AAs, they may need to wait for a certain time delay. This delay is caused by two factors: 1) the round-trip time required to get the commitments and signatures, and 2) the lack of incentive for the AAs to sign off the content quickly. However this risk is mitigated by the fact that the CC can initiate the sign off request at any point of time, even after the content has been published. In addition, they can publish their content (along with the signatures they received so far), and then they can initiate another request to get more signatures on the content.

- Once the collective signature along with the content hash has been added in the blockchain, the AAs cannot revoke their decision because there is no way to revert that addition in blockchain. However, we are planning to implement a mechanism to allow any AA to retrigger the sign off process which will result in a new collective signature.

\subsection{Advantages of the TRUSTD approach}
TRUSTD approach can be applied not only for combating the fake news, but it could be extended to other domains as well. For example, a new type of a decentralised social network can be built in which CCs and AAs help each other. Contrast to the existing centralised social media networks, a decentralised network can be built where CCs produce different types of content and get different types of appraisal such as "Agree" (I agree with the point of view of the content), "Correct" (the content is correct), "Support" (I support the call of the content), etc. (in form of signature) from different individuals and organisations without any central entity governing the whole network. For example, authors can get their research conference/journal papers  marked as "correct" by the reviewers, bloggers can get their opinion blogs marked as "agreed upon" by a set of authors/people, petition creators can get their petitions to request some change marked as "supported" by a set of people, etc. In all cases, users are able to judge on the credibility of any content based on their trust policy.

\section{Conclusion}
We look at spreading the fake news as a special case of spreading all kinds of deceptive contents. The use of online social media to connect with people around the world is rising sharply and it has become an important source for the distribution of digital deception. This setting shifts the responsibility of verifying the content to these centralised social media platforms. In addition, the user has no say or control over the filtering mechanism implemented by these platforms. TRUSTD allows users to specifies their trusted parties and to determine the credibility level of any published content. We believe this is a step in the right direction where the ecosystem is decentralised, all users, AAs and CCs own their identities, and the user is in the center of the equation. 

\bibliographystyle{IEEEtran}
\bibliography{references}

\begin{thebibliography}{10}
\providecommand{\url}[1]{#1}
\csname url@samestyle\endcsname
\providecommand{\newblock}{\relax}
\providecommand{\bibinfo}[2]{#2}
\providecommand{\BIBentrySTDinterwordspacing}{\spaceskip=0pt\relax}
\providecommand{\BIBentryALTinterwordstretchfactor}{4}
\providecommand{\BIBentryALTinterwordspacing}{\spaceskip=\fontdimen2\font plus
\BIBentryALTinterwordstretchfactor\fontdimen3\font minus
  \fontdimen4\font\relax}
\providecommand{\BIBforeignlanguage}[2]{{%
\expandafter\ifx\csname l@#1\endcsname\relax
\typeout{** WARNING: IEEEtran.bst: No hyphenation pattern has been}%
\typeout{** loaded for the language `#1'. Using the pattern for}%
\typeout{** the default language instead.}%
\else
\language=\csname l@#1\endcsname
\fi
#2}}
\providecommand{\BIBdecl}{\relax}
\BIBdecl

\bibitem{survey2}
X.~Zhou and R.~Zafarani, ``Fake news: A survey of research, detection methods,
  and opportunities,'' vol.~1, no.~1, 2018.

\bibitem{james}
J.~Mahon, ``Kant on lies, candour and reticence,'' \emph{Kantian Review},
  vol.~7, 03 2003.

\bibitem{oncombating}
T.~Zubair, A.~Raquib, and J.~Qadir, ``On combating fake news, misinformation,
  and machine learning generated fakes: Insights from the islamic ethical
  tradition,'' 10 2019.

\bibitem{bubble}
D.~Difranzo and K.~Gloria-Garcia, ``Filter bubbles and fake news,'' \emph{XRDS:
  Crossroads, The ACM Magazine for Students}, vol.~23, pp. 32--35, 04 2017.

\bibitem{did}
\BIBentryALTinterwordspacing
``Decentralized identifiers (dids) v1.0, w3c working draft,'' 2019. [Online].
  Available: \url{https://www.w3.org/TR/did-core/}
\BIBentrySTDinterwordspacing

\bibitem{syta2016keeping}
E.~Syta, I.~Tamas, D.~Visher, D.~I. Wolinsky, P.~Jovanovic, L.~Gasser,
  N.~Gailly, I.~Khoffi, and B.~Ford, ``Keeping authorities" honest or bust"
  with decentralized witness cosigning,'' in \emph{2016 IEEE Symposium on
  Security and Privacy (SP)}.\hskip 1em plus 0.5em minus 0.4em\relax Ieee,
  2016, pp. 526--545.

\bibitem{survey1}
G.~Haciyakupoglu, J.~Y. Hui, V.~Suguna, D.~Leong, and M.~F. B.~A. Rahman,
  ``Countering fake news: A survey of recent global initiatives,'' 2018.

\bibitem{facebook1}
A.~Guess, J.~Nagler, and J.~Tucker, ``Less than you think: Prevalence and
  predictors of fake news dissemination on facebook,'' \emph{Science Advances},
  vol.~5, p. eaau4586, 01 2019.

\bibitem{google}
``How google fights disinformation,'' 2019.

\bibitem{ieee-initative}
\BIBentryALTinterwordspacing
``Ieee global initiative et al., ethically aligned design, ieee standards v1,''
  2016. [Online]. Available:
  \url{https://standards.ieee.org/content/dam/ieee-standards/standards/web/documents/other/ead_v2.pdf}
\BIBentrySTDinterwordspacing

\bibitem{Shang}
W.~Shang, M.~Liu, W.~Lin, and M.~Jia, ``Tracing the source of news based on
  blockchain,'' 06 2018, pp. 377--381.

\bibitem{bc1}
S.~Huckle and M.~White, ``Fake news: A technological approach to proving the
  origins of content, using blockchains,'' \emph{Big Data}, vol.~5, pp.
  356--371, 12 2017.

\bibitem{ohta1999multi}
K.~Ohta and T.~Okamoto, ``Multi-signature schemes secure against active insider
  attacks,'' \emph{IEICE Transactions on Fundamentals of Electronics,
  Communications and Computer Sciences}, vol.~82, no.~1, pp. 21--31, 1999.

\bibitem{cosi-ietf}
B.~Ford, N.~Gailly, L.~Gasser, and P.~Jovanovic, ``Collective edwards-curve
  digital signature algorithm,'' Internet-Draft draft-ford-cfrg-cosi-00. txt.
  IETF Secretariat, Tech. Rep., 2017.

\bibitem{ssi}
A.~Mühle, A.~Grüner, T.~Gayvoronskaya, and C.~Meinel, ``A survey on essential
  components of a self-sovereign identity,'' 07 2018.

\bibitem{vc}
\BIBentryALTinterwordspacing
``Verifiable claims working group frequently asked questions,'' 2019. [Online].
  Available: \url{https://w3c.github.io/webpayments-ig/VCTF/charter/faq.html}
\BIBentrySTDinterwordspacing

\bibitem{did_spec}
\BIBentryALTinterwordspacing
``Decentralized {Identifiers} ({DIDs}) v0.11.'' [Online]. Available:
  \url{https://w3c-ccg.github.io/did-spec/}
\BIBentrySTDinterwordspacing

\bibitem{bc2}
A.~Shahaab, B.~Lidgey, C.~Hewage, and I.~Khan, ``Applicability and
  appropriateness of distributed ledgers consensus protocols in public and
  private sectors: A systematic review,'' \emph{IEEE Access}, vol.~PP, pp.
  1--1, 03 2019.

\bibitem{bc3}
P.~Fraga-Lamas and T.~Fernández-Caramés, ``A review on blockchain
  technologies for an advanced and cyber-resilient automotive industry,''
  \emph{IEEE Access}, vol.~7, pp. 17\,578--17\,598, 01 2019.

\bibitem{bc4}
T.~M. Fernández-Caramés and P.~Fraga-Lamas, ``A review on the application of
  blockchain for the next generation of cybersecure industry 4.0 smart
  factories,'' 2019.

\bibitem{OnTheSecurity}
M.~Drijvers, K.~Edalatnejad, B.~Ford, E.~Kiltz, J.~Loss, G.~Neven, and
  I.~Stepanovs, ``On the security of two-round multi-signatures,'' 05 2019, pp.
  1084--1101.

\end{thebibliography}

\end{document}